\documentclass[a4paper]{article}
\usepackage[utf8]{inputenc}
\usepackage[T1]{fontenc}
\usepackage[english]{babel}
\usepackage{amsmath,amsfonts,graphicx,mathptmx}

\date  {}
\title {Clustering Macroeconomic Time Series\texorpdfstring{\footnote{Published in
  \emph{Econometrics. Ekonometria. Advances in Applied Data Analysis},
  22 (2018) 2, pp.\ 74--88, DOI: 10.15611/eada.2018.2.06}}{}}
\author{%
  Iwo Augustyński\texorpdfstring{%
    \thanks{Wrocław University of Economics, Wrocław, Poland; e-mail:
    \texttt{\href{mailto:iwo.augustynski@ue.wroc.pl}{iwo.augustynski@ue.wroc.pl}}}
  \and}{,}%
  Paweł Laskoś-Grabowski\texorpdfstring{%
    \thanks{Institute of Theoretical Physics, University of Wrocław, Poland}}{}}

\newcommand \keywords {time series clustering, similarity, cluster analysis, GDP}
\newcommand \jelcodes {E00, C18, C63}

\usepackage[unicode=true,pdftex]{hyperref}
\makeatletter
\hypersetup{pdftitle={\@title},pdfauthor={\@author},pdfkeywords={\keywords},pdfsubject={\jelcodes}}
\makeatother

\usepackage[apaciteclassic]{apacite}
\begin{document}
\renewcommand \BBAB \&
\def\sectionautorefname{Section}
\maketitle
\renewcommand \thefootnote {\fnsymbol{footnote}}

\begin{abstract}
The data mining technique of time series clustering is well established in many fields. However, as an unsupervised learning method, it requires making choices that are nontrivially influenced by the nature of the data involved.
The aim of this paper is to verify usefulness of the time series clustering method for macroeconomics research, and to develop the most suitable methodology.

By extensively testing various possibilities, we arrive at a choice of a dissimilarity measure (compression-based dissimilarity measure, or CDM) which is particularly suitable for clustering macroeconomic variables. We check that the results are stable in time and reflect large-scale phenomena such as crises. We also successfully apply our findings to analysis of national economies, specifically to identifying their structural relations.
\\[\baselineskip]\textbf{JEL}: \jelcodes
\\[\baselineskip]\textbf{Keywords}: \keywords
\end{abstract}

\section{Introduction}

The algorithms for clustering similar time series or, more generally, similar high-dimensional sequences, are important in areas as diverse as biomedicine, computational biology, electronic manufacturing, physics, seismology and speech recognition.

Econometrics could also benefit from a vast research effort made in these and other areas. For example, according to \shortciteA{focardi_clustering_2004}, clustering of economic and financial time series includes the following areas of application:
\begin{itemize}
\item identifying areas or sectors for policy-making purposes;
\item identifying structural similarities in economic processes for economic forecasting;
\item identifying stable dependencies for risk management and investment management.
\end{itemize}

For studies in macroeconomics, one of the most promising advantages of time series clustering is its ability to identify structural similarities in the processes that generate time series at different points in time and space. This method also allows for presentation of results in the form of easy-to-understand dendrograms.

The aim of this paper is to find the most appropriate dissimilarity measure for macroeconomic clustering analysis.

The problem of time series similarity could be boiled down to measurement of the co-movement of macroeconomic aggregates. The most popular approaches to this issue are \shortcite{croux_measure_2001,haan_will_2005}:
\begin{itemize}
\item correlation;
\item cointegration, that is, the existence of a linear combination of the two processes that is stationary;
\item codependence, which refers to linear combinations of correlated processes that are of lower autoregressive order than others;
\item common features, that is, linear combinations that are unpredictable with respect to past information, and common cycles which are defined as common features in first differences for processes that are cointegrated.
\end{itemize}

According to \shortciteA{croux_measure_2001}, these concepts pose several problems. First, correlations might be detected where no correlation is present. Second, high cross-correlation neither implies nor is implied by cointegration, common cycles, or common features. Third, these three measures are binary. For example, two processes are either cointegrated or not, but different degrees of association cannot be established. Tests such as the Johansen test, performed by commercial econometric packages, consist of fitting empirical data to cointegrated models such as Error Correcting Models (ECM). In the case of a large number of time series, cointegration is a rather cumbersome exploratory methodology.

The contribution of the present paper to the literature is as follows:
\begin{itemize}
\item to our knowledge, it is the first comprehensive analysis of the usefulness of the different dissimilarity measures for the macroeconomic research,
\item it offers ready-to-use methodology,
\item it offers tools to present data in easy-to-understand dendrograms,
\item it is provided with code and web application for easy use\footnote{\url{http://iwoaugustynski.ue.wroc.pl/apps/TSclustering/}}.
\end{itemize}

The remainder of this paper is organized as follows: in \autoref{s:exp} we evaluate available dissimilarity measures and propose CDM (compression-based dissimilarity measure) as a solution for clustering of the macroeconomic time series. In \autoref{s:app} we check the robustness of the presented methodology by applying the proposed clustering method and comparing created clusters with the literature. Finally, in \autoref{s:fin} we present concluding remarks.

All figures are the results of own calculations based on R package TSclust and Eurostat data (\texttt{namq\_10\_gdp} dataset).

\section{Experimental Evaluation of Dissimilarity Measures}\label{s:exp}

A well-known data mining technique, clustering is an example of \emph{unsupervised} learning: a clustering algorithm creates clusters as a function of its internal rules (whereas in supervised learning the algorithm learns from “known” examples). The objective of clustering is to create groups of objects that are close to each other and distant from other groups of objects. If distance corresponds to similarity, clustering forms groups of objects that are maximally similar.

\subsection{Design of the Experimental Approach}

A dissimilarity measure suitable for our applications could be informally described as attaining low values for pairs of time series that exhibit a causal relationship. This property is hard, if not outright impossible, to formulate in rigorous terms, but it can be approximated well enough by demanding that the measure is insensitive to \emph{translating} the series in time. Importantly, it \emph{should} be sensitive to other time transforms, including warping (acceleration), whether uniform or not. Conversely, other kinds of transforms, such as scaling the values by a constant, adding a constant to the values, or adding noise to the values, should not, in principle, affect the measure much.

We can apply this observation to design an experiment to identify prospective measures. For a given time series, we can compute the distance separating it from its delayed copy (i.e.\ a series with the same data, but shifted in time), as well as the distance separating it from its warped copy (i.e.\ a series with the same data, but squeezed in time). The ratio of these distances is a measure of how well a dissimilarity measure performs. We are looking for a measure that performs well for as many as possible time series, delays, and warp factors. Furthermore, this should hold true even if the delayed copy of the series is additionally subjected to \emph{perturbations} such as scaling, shifting, noise, or a combination thereof.

To formalize the idea, we fix a time series $X=\{X_i\}_{i=1}^T$, dissimilarity measure $M$, delay $\delta\in\mathbb{N}$, and warp factor $\alpha\in\mathbb{R}$. We call the subseries $B[X;\alpha]=\{X_i\}_{i=1}^{\lfloor T/\alpha\rfloor}$ and $D[X;\alpha,\delta]=\{X_i\}_{i=1+\delta}^{\lfloor T/\alpha\rfloor+\delta}$ respectively the \emph{base} and \emph{delayed} series. We also obtain from $X$ the \emph{warped} series $W[X;\alpha]=\{\bar X^{(\alpha)}_i\}_{i=1}^{\lfloor T/\alpha\rfloor}$ by taking averages of every $\alpha$ consecutive elements, trivially extending the notion for non-integer $\alpha$. Strictly speaking, $\bar X^{(\alpha)}_i=\frac1\alpha\sum_{j=1}^T w_{ij}^{(\alpha)} X_j$, where
\begin{align*}
w_{ij}^{(\alpha)}=\left\{
\begin{array}{c@{}c@{\quad\text{if}\quad}c@{\;<\;}c@{\;<\;}c@{\;<\;}c}
\alpha        && j-1         & (i-1)\alpha & i\alpha & j       \\
j-(i-1)\alpha && j-1         & (i-1)\alpha & j       & i\alpha \\
1             && (i-1)\alpha & j-1         & j       & i\alpha \\
i\alpha-(j-1) && (i-1)\alpha & j-1         & i\alpha & j       \\
0             & \multicolumn{5}{@{}l}{\quad\text{otherwise.}}  \\
\end{array}
\right.
\end{align*}

Let $d_M(\cdot,\cdot)$ be the distance between two time series under the measure $M$. We will use the ratio $R(M,X,\delta,\alpha)=\frac{d_M(B[X;\alpha],\tilde D[X;\alpha,\delta])}{d_M(B[X;\alpha],W[X;\alpha])}$ as a quality indicator, where $\tilde D[X;\alpha,\delta]$ is $D[X;\alpha,\delta]$ with perturbation applied. Note that $X$ is truncated to its prefix $B[X;\alpha]$ for calculating $R$ to fulfill a requirement of many of the measures that the series compared be of the same length.

Note that while the values of distances under different dissimilarity measures are not directly comparable to one another (as different measures may e.g.\ attain values in different ranges), ratios such as $R$ are. Note also that if a measure is \emph{good}, that is, in general it separates warped series more than the delayed series, its values of $R$ will be in general (or, ideally, always) less than 1. Note finally that a dissimilarity measure is \emph{better} than another if its values of $R$ are in general smaller.

\subsection{Sampled Measures, Series, Parameters, and Perturbations}

In our computations, we consider all dissimilarity measures provided by the R package TSdist, which, for the sake of simplicity and avoiding potential cognitive bias, may be calculated without supplying extra parameters. This is possible either thanks to their absence, or default values, or heuristics. These 24 measures are (referred to by names used within TSdist): \texttt{euclidean}, \texttt{manhattan}, \texttt{infnorm}, \texttt{ccor}, \texttt{sts}, \texttt{dtw}, \texttt{fourier}, \texttt{acf}, \texttt{pacf}, \texttt{ar.lpc.ceps}, \texttt{ar.mah.statistic}, \texttt{ar.pic}, \texttt{cdm}, \texttt{cid}, \texttt{cor}, \texttt{cort}, \texttt{wav}, \texttt{int.per}, \texttt{per}, \texttt{ncd}, \texttt{spec.glk}, \texttt{spec.isd}, \texttt{spec.llr}, and \texttt{pdc}.

Furthermore, we consider warp factors $\alpha$ between 1.4 and 3.0 inclusive in steps of 0.2, and delays $\Delta$ of 2 through 10 quarters. Note that for series with quarterly data, the delay in terms of data points is $\delta=\Delta$, while for monthly data it is $\delta=3\Delta$.

We also consider \emph{separately} two different sets of time series, along with perturbations specific for each set. The first set contains 52 time series of \emph{absolute-valued} data:
\begin{itemize}
\item Eurostat quarterly GDP values for 28 EU Member States,
\item Eurostat quarterly GDP component values for 11 components of the UK GDP (with total GDP “component” omitted, as it is present in the preceding category),
\item three series obtained from the FRED quarterly UK GDP values by reversing the sign as well as concatenating prefixes and suffixes of this and the original series,
\item FRED monthly long-term (ten-year) government bond yields for USA, Germany, and France,
\item FRED monthly short-term (three-month) certificates of deposit yields for USA and interbank rates for Germany and France,
\item four artificial series: sine and triangular waves of three periods (each of 100 “months” for the sake of compatibility), with either constant or linearly diverging extrema.
\end{itemize}

The following perturbations were applied to the delayed copies of the series from this set in distinct computation runs:
\begin{itemize}
\item multiplying the values by a constant,
\item adding a constant (proportional to the standard deviation of the original series) to the values,
\item adding random noise (also proportional to the standard deviation) to the values,
\item all of the above,
\item none of the above.
\end{itemize}

The second set contains 45 time series of \emph{quarter-to-quarter percentage changes} of the same quantities as in the first set, except the three artificial modifications of the UK GDP, as well as four of the UK GDP components for which the percentage data is not available (i.e.\ compensation of employees, taxes on production and imports less subsidies, changes in inventories and acquisitions less disposals of valuables, and operating surplus and mixed income, gross). Also, the four artificial (sine and triangular wave) series are not converted to percentage changes, but (with different value ranges) treated as percentage change series in their own right. Given the different nature of the data in this set, we deem the perturbations of scaling and shifting inapplicable here, and computations are performed in two runs (without perturbations and with random noise, not adjusted for standard deviation in this case).

In each of the seven runs, ratios $R$ are computed for all possible combinations of measures, warp factors, delays, and time series.

\subsection{Evaluation Results}

Examining the results of all seven computation runs, we have primarily focused on the following quantities:
\begin{itemize}
\item $\max_{X,\Delta,\alpha} R$ for every measure $M$, that is, the maximum value of ratio $R$ achieved across all time series, delays, and warp factors;
\item the count of how often (out of 81 possible combinations of delay $\Delta$ and warp factor $\alpha$) does measure $M$ rank either first or in top five (of 24 measures considered) when ordered by $\max_X R$ (for given $M,\Delta,\alpha$).
\end{itemize}

Based on this, we arrive at the conclusion that \texttt{cdm} is the overall best performing measure. Specifically (see also \autoref{t:standings}):

\begin{itemize}
\item Without perturbations, both for absolute- and percentage-valued series, \texttt{cdm} is the only measure except for \texttt{ncd} that never exceeds 1 (and thus is \emph{good} in the sense outlined above). Note that in these runs \texttt{ncd} performs better than \texttt{cdm}, with lower global maximum and more frequent appearances in the top spot or top five spots. In fact, \texttt{cdm} never ranks first in these cases, although only three measures ever do (\texttt{ncd}, \texttt{dtw}, and, only once for absolute- and five out of 81 times for percentage-valued series, \texttt{pdc}), and ranks third most often in the top five spots (behind \texttt{ncd} and \texttt{dtw} again).

\item In the remaining five runs, \texttt{cdm} has the lowest global maximum, which additionally is always less than ca. $\frac43$. For two runs (absolute-valued series with scaling and with all perturbations at once), it is the only measure with a maximum less than 2, while additionally for absolute-valued series with shifting \texttt{ncd} is the only other measure with a maximum less than 2.

\item For absolute-valued series with all perturbations at once, \texttt{cdm} ranks most often in the top spot, and, importantly, it does so for more than half of the possible combinations of warp factor and delay. Additionally, it ranks second most often in the top spot for absolute-valued series with scaling and with shifting (behind \texttt{per}), and for absolute- and percentage-valued series with noise (behind \texttt{dtw}).

\item For four runs (absolute-valued series with scaling, with shifting, and with all perturbations at once, as well as percentage-valued series with noise), \texttt{cdm} ranks most often in top five spots (although on par with \texttt{pdc} for absolute-valued series with shifting). For absolute-valued series with scaling and with all perturbations at once, it always ranks in top five spots, while for percentage-valued series with noise, it fails to do so for only three combinations of warp factor and delay. For absolute-valued series with noise, \texttt{cdm} ranks third most often in top five spots (behind \texttt{dtw} and \texttt{cid}).
\end{itemize}

Another criterion that could be sensibly used here is the count of how often the ratios for a given measure exceed 1. Admittedly, with perturbations present, \texttt{cdm} does not perform well in this aspect. However, precisely in the presence of perturbations this requirement can be argued to be too restrictive: the delayed \emph{and} perturbed series can be informally considered as at a disadvantage compared to its warped and \emph{not} perturbed counterpart (especially if perturbations are, in some sense, large). It also does not give consistent conclusions across different perturbations: \texttt{pdc} ranks best for absolute-valued series with scaling or shifting, \texttt{dtw} for absolute- and percentage-valued series with noise, and \texttt{int.per} for absolute-valued series with all perturbations at once. In general, for absolute-valued series, measures performing well with noise mostly perform poorly with other two perturbations, and vice versa. Finally, measures should perform well also without perturbations, and that in these cases \texttt{cdm} together with \texttt{ncd} rank best as the only \emph{good} measures.

CDM (compression-based dissimilarity measure) introduced by \shortciteA{keogh_towards_2004} and further elaborated upon in \shortciteA{keogh_compression-based_2007}, has already been demonstrated (e.g.\ in these two references) to be immensely useful in various data mining aspects, prominently including time series clustering. It warrants mention, however, that it is not a \emph{distance measure} (which is why we avoid using that phrase altogether throughout this paper), as, among other properties, by definition it attains values in the range $[\frac12,1]$, i.e.\ does not reach 0, even for equal arguments. This also means that the ratios $R$ we compute in our experiment are always going to be within the range $[\frac12,2]$ for CDM, and it may be argued that the setup is biased in favour of this measure. However, we uphold our conclusion because no other measure emerges as a clear alternative, especially if we demand that it performs well both with noise and with other types of perturbations. As a bonus, CDM also performs well with both the absolute- and percentage-valued series, making this measure much more versatile, even though we do not explicitly require such behaviour. Also, the unambiguously good results without perturbations (i.e.\ ratios below 1) and aforementioned reports of CDM suitability for clustering in general support this choice.

\subsection{Choice of Clustering Method}

The next step in the procedure was to choose the most suitable clustering method. After clustering one should obtain a figure resembling a tree with some main branches and many smaller side branches. Conversely, a structure of ascending steps would disqualify a given clustering method.

We tested the following approaches:
\begin{itemize}
\item single linkage,
\item complete linkage,
\item Ward,
\item average (UPGMA),
\item McQuitty (WPGMA),
\item median (WPGMC),
\item centroid (UPGMC).
\end{itemize}

Only the first five of these produced useful dendrograms. Ultimately we decided to pick the Ward method, as it is less sensitive to changes in the length of the time series and creates better separated groups.

\section{Empirical Application}\label{s:app}

For the purposes of this paper we employed the time series of GDP for analysis of the measure's stability, and its direct components (investment, consumption, import, export, employment, wages, etc.) to compare the economic structures of four EU Member States: Germany, France, Italy, and Spain. The reasoning for this choice is as follows: all of them are members of the Eurozone, Germany and France are the most advanced European countries, whereas Italy and Spain are the biggest lagging-behind members of the EU.

In order to establish international comparisons, we used data from the Eurostat after the introduction of the euro currency, that is, from the first quarter of 2000.

We decided to use the “raw” data, that is, quarterly, not seasonally adjusted time series. We argue that using data subject to preprocessing such as seasonal adjusting or detrending could introduce artificial distortions to the time series \shortcite{hamilton_why_2017,haan_will_2005}.

\subsection{Type of Data}

In the following calculations we used nominal GDP in millions of euro, but other options are also available:
\begin{itemize}
\item real (chain linked) values,
\item national currency,
\item first differences (or percentage change).
\end{itemize}

Which one is the most suitable for the considered method?

First, it depends on data availability. Usually the most up-to-date and the most accessible data is represented in nominal values in a given national currency.

Second, when comparing the behaviour of different real variables we are usually not interested in their nominal values, but their changes. Therefore, percentage change is the one most commonly used. But it is problematic in the case of, for instance, financial data, which is often presented as a percent rate (interest rates, yields, etc.). Percentage change of a percent rate may be hard to interpret.

Third, the answer to the question if real or nominal values are preferable is much more straightforward. If the dataset contains variables measured both in real terms (such as the number of unemployed) and in currency terms (such as consumption or GDP), chain-linked values should be employed. Otherwise, when all variables are represented in currency terms, distortions caused by inflation could be probably neglected.

We argue that the proposed dissimilarity measure is suitable in all of the aforementioned cases.

\subsection{Clustering Stability}

\begin{figure}[p!]
\centering
\includegraphics[width=0.75\textwidth]{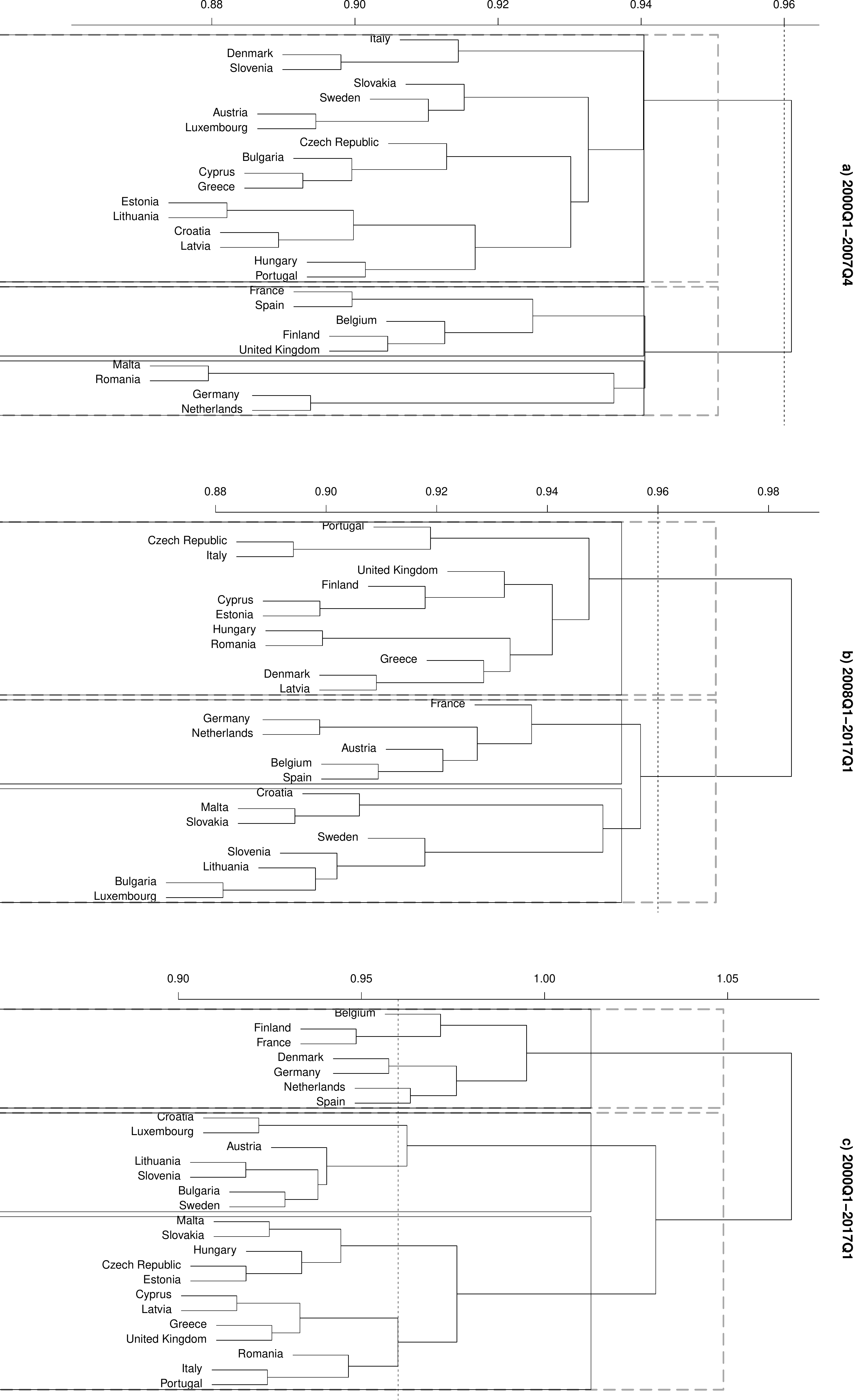}
\caption{GDP time series clustering. Quarterly data for the EU Member States (Poland and Ireland excluded because of the data availability), millions of euro.\label{f:countries}}
\end{figure}

We begin empirical testing with analysis of the time stability of the CDM measure. To this end, we compare time series of the GDPs of the EU Member States, covering three periods: 2000Q1–2007Q4, 2008Q1–2017Q1, and the complete period of 2000Q1–2017Q1 (\autoref{f:countries}).

The first conclusion is rather obvious: the longer the time series are, the less similarity emerges. This phenomenon is indicated by the dotted line.

A related remark is that the financial crisis reversed processes of economic integration within the EU (both periods are of similar length). This is consistent with the findings of \shortciteA{belke_business_2017,gachter_business_2012,ahlborn_output_2017}, who employed different synchronization measures (respectively: correlation, panel regressions, and nonparametric regressions; correlation; and fuzzy clustering).

The employed similarity method allows us to distinguish two main groups of countries (grey outer rectangles in \hyperref[f:countries]{Figure \ref*{f:countries}c}): the “core”, consisting of France, Germany, the Netherlands, Austria, Belgium, and Spain; and the “periphery”. This structure was slightly different before and after the financial crisis (respectively, \hyperref[f:countries]{Figures \ref*{f:countries}a and \ref*{f:countries}b}).

Hierarchical representation of the distances enables changing the level of clustering (black inner rectangles). On this lower level, in the first period (2000–2007) the countries formed four quite similar groups consisting of: (a) Italy, Denmark, and Slovenia; (b) Germany, the Netherlands, Malta, and Romania; (c) France, Spain, UK, Finland, and Belgium; (d) the remaining 14 Member States.

In the second period, generally speaking, after the financial crisis, countries are less similar and it is more appropriate to cluster them into three groups: (a) the “core”, consisting of France, Germany, the Netherlands, Austria, Belgium, and Spain; (b) the “semi-periphery”, consisting of Croatia, Malta, Slovakia, Sweden, Slovenia, Lithuania, Bulgaria, and Luxembourg; (c) the “periphery”, consisting of the remaining 12 countries, including Greece, Italy, Portugal, and the United Kingdom. This structure holds also for the full time series.

These results are similar to (countries in common typeset in boldface):
\begin{itemize}
\item \shortciteA{belke_business_2017}, who distinguished \textbf{Finland}, \textbf{France}, \textbf{Germany}, Austria, and \textbf{the Netherlands} as the core countries;
\item \shortciteA{papageorgiou_business_2010}, where the core countries group in 2000–2009 consists of Sweden, Portugal, \textbf{Germany}, \textbf{France}, \textbf{Spain}, \textbf{Belgium}, Denmark, Austria, and \textbf{the Netherlands};
\item \shortciteA{ahlborn_output_2017}, who grouped Austria, \textbf{Belgium}, \textbf{Denmark}, \textbf{Finland}, \textbf{France}, \textbf{Germany}, Hungary, Ireland, \textbf{the Netherlands}, Norway, Poland, Portugal, \textbf{Spain}, Sweden, Switzerland, and the United Kingdom as the core countries in 1996Q1–2015Q4.
\end{itemize}

To validate the usefulness of the proposed method in the analysis of the structure of an economy, we have selected variables used to calculate the gross domestic product (GDP) and GDP itself.

\subsection{Analysis of National Economies}

According to \shortciteA[p.\ 273]{esa_2010_european_2013}, there are three approaches to calculating GDP:
\begin{description}
\item[production approach] GDP is the sum of the gross value added of the various institutional sectors or the various industries plus taxes and minus subsidies on products (which are not allocated to sectors and industries);
\item[expenditure approach] GDP is the sum of final uses of goods and services by resident institutional units (final consumption of general government, households, and NPISH and gross capital formation) plus exports and minus imports of goods and services;
\item[income approach] GDP is the sum of uses in the total economy generation of income account (compensation of employees, taxes on production and imports minus subsidies, gross operating surplus and mixed income of the total economy).
\end{description}

Taking into consideration economic theory and how GDP is calculated, the following distances between variables should be small in all of the analysed countries:
\begin{itemize}
\item \emph{GDP} and \emph{gross value added},
\item \emph{compensation of employees} and \emph{household and NPISH final consumption expenditure}.
\end{itemize}

\begin{figure}[t]
\centering
\includegraphics[width=\textwidth]{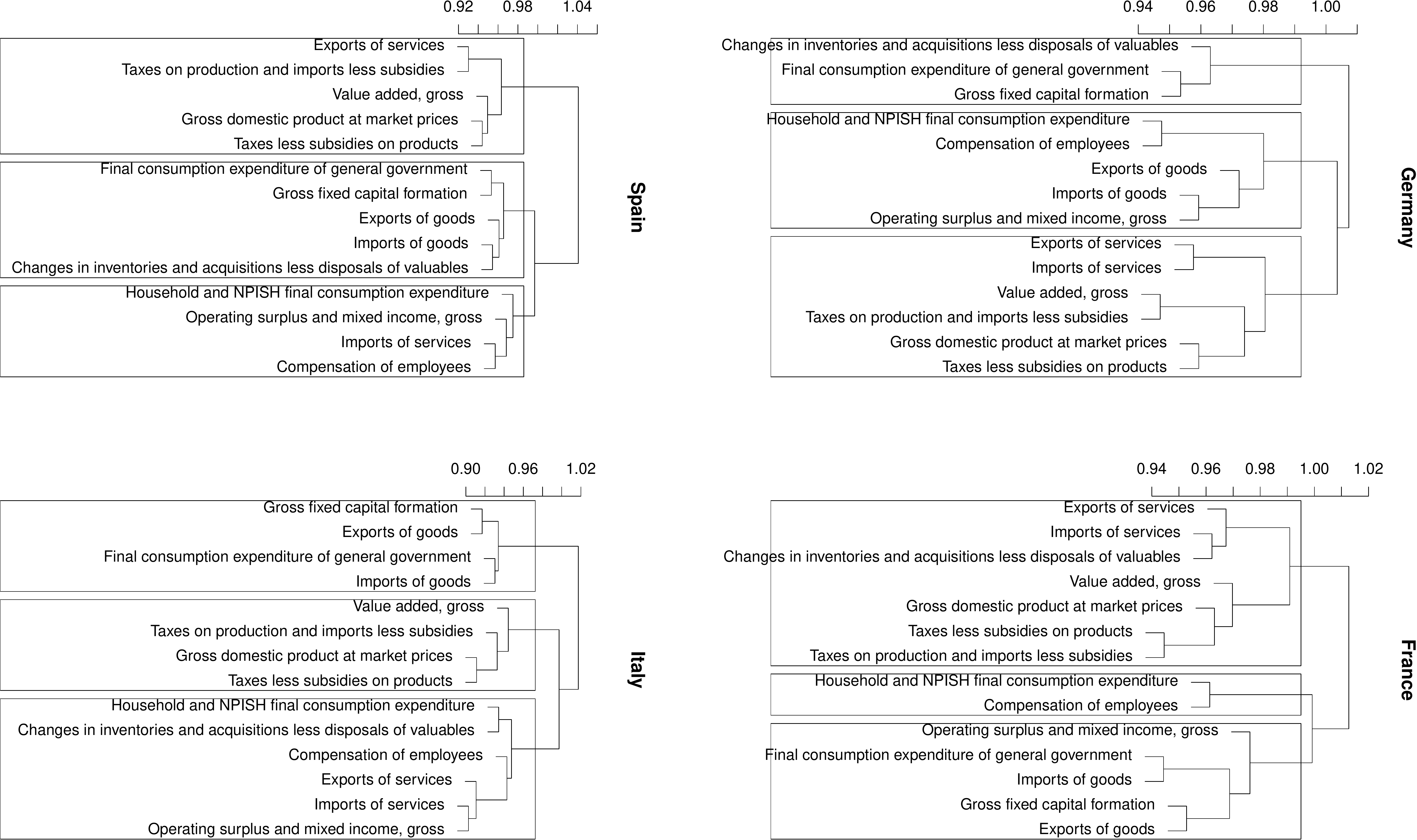}
\caption{Structure of GDP and its components in Germany, France, Italy, and Spain. 2000Q1–2017Q1 period, millions of euro.\label{f:components}}
\end{figure}

To verify these relations, we performed an agglomerative clustering on a set of 14 macroeconomic variables (for a full list see \autoref{f:components}) that include these four, for each of the 28 EU Member States in the period of 2002Q1–2016Q4. The computation begins with all the variables in separate clusters, and gradually links them together until the two variables of the pair are in one cluster. The number of clusters at this stage is indicative of how similar are the variables of the pair in comparison with the others: the earlier they are clustered together, or the higher the final number of clusters, the relatively more similar they are.

\begin{figure}[t]
\centering
\includegraphics[width=\textwidth]{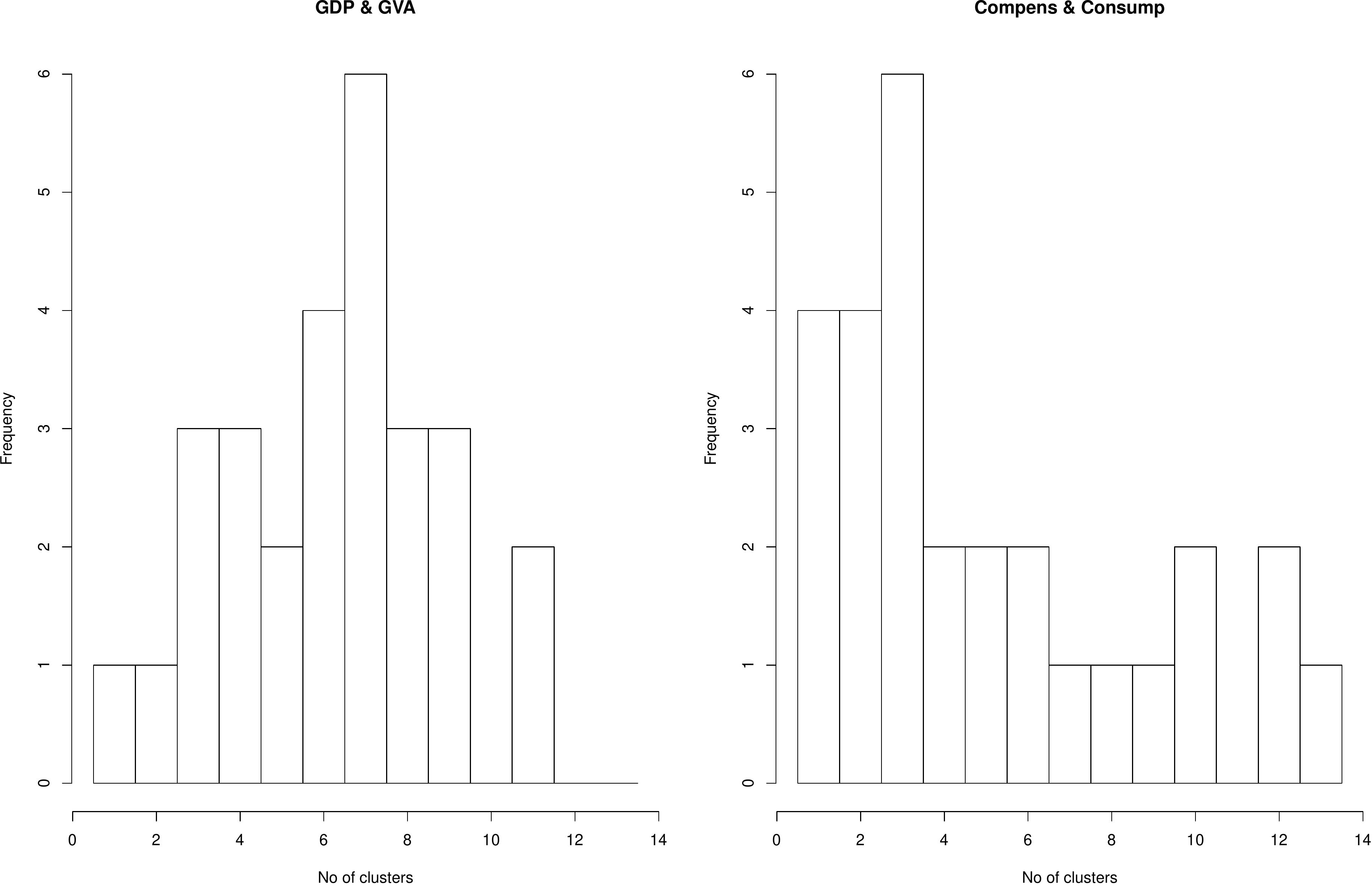}
\caption{Histograms of clustering stages at which (left) \emph{GDP} and \emph{gross value added}, or (right) \emph{compensation of employees} and \emph{household and NPISH final consumption expenditures} become clustered together for each of the 28 EU Member States.\label{f:histogram}}
\end{figure}

As a rule of thumb, we propose that if compared variables belong to the same group when three or more clusters are created, then the method proposed in this paper is positively verified (confront \hyperref[f:histogram]{Figures \ref*{f:histogram}} and \ref{f:components}).

In the case of \emph{GDP} and \emph{gross value added}, there were only two countries for which these variables were linked only in the two last steps of cluster merging. Note that such a situation should raise questions about data quality.

In the case of \emph{compensation of employees} and \emph{household and NPISH final consumption expenditures}, the relation is not so straightforward, and therefore the relative distances are larger. Nevertheless, in half of the analysed countries these variables were connected when the number of clusters was more than three.

In our opinion, these results confirm that the proposed method is consistent with SNA methodology and economic theory.

Finally, we selected four countries (Germany, France, Italy, and Spain) for more detailed verification. Relations between variables are compared in \autoref{f:components}. As expected, every country has its unique structure.

Nevertheless, the macroeconomic indicators in Spain, Germany, and Italy form three main clusters, while in the case of France it is more natural to distinguish four main groups.

A more detailed look into the selected countries reveals that all three approaches to the calculation of GDP are generally preserved. This also confirms the robustness of the proposed method.

\section{Concluding Remarks}\label{s:fin}

The aim of this paper was to verify usefulness of the time series clustering method for macroeconomics research, and to develop the most suitable methodology.

After the evaluation of 24 dissimilarity measures, the CDM measure was assessed as the most suitable for our purposes. It most effectively prefers time series with similar frequencies over similar magnitudes of volatility. We assume that likeliness of frequencies could suggest causality.

This assumption and the usefulness of the proposed method is confirmed in the analysis of the distances between aggregated time series of the EU Member States. Results confirm that CDM preserves relations from the national accounts equations as well as from economic theory. It is also robust for manipulations in the length of time series.

It could be therefore recommended for a variety of macroeconomic research topics, including, but not limited to, issues of balance of payments, fiscal and monetary policy, or financial integration in the EU.

\bibliography{clustering}

\renewcommand \baselinestretch 1

\begin{table}[t]
\begin{center}
\begin{tabular}{c @{\qquad} lr@{.}l @{\qquad} r@{}l@{}r@{}l @{\qquad} lr}\hline

Value type,  & \multicolumn{3}{@{}c@{\qquad}}{Global}  & \multicolumn{6}{c}{Number of times ranked} \\\cline{5-10}
perturbation & \multicolumn{3}{@{}c@{\qquad}}{maximum} & & \multicolumn{2}{c}{first}        &     & \multicolumn{2}{c}{top five}  \\\hline\hline

Absolute     & \texttt{ncd}                     & 0&66 & & \texttt{ncd}                & 68 &     & \texttt{ncd}             & 81 \\
None         & \underline{\texttt{cdm}}         & 0&85 & & \texttt{dtw}                & 12 &     & \texttt{dtw}             & 72 \\
             & \texttt{dtw}                     & 1&44 & & \texttt{pdc}                &  1 &     & \underline{\texttt{cdm}} & 62 \\
             &              & \multicolumn{1}{r}{} &   &(& \underline{\texttt{cdm}}    &  0 & )   & \texttt{pdc}             & 50 \\\hline
         
Absolute     & \underline{\texttt        {cdm}} & 1&33 & & \texttt{pdc}                & 50 &     & \underline{\texttt{cdm}} & 81 \\
Scaling      & \texttt{ncd}                     & 2&07 & & \underline{\texttt{cdm}}    & 22 &     & \texttt{ncd}             & 74 \\
             & \texttt{cor}                     & 2&20 & & \texttt{int.per}            &  8 &     & \texttt{pdc}             & 69 \\
             &              & \multicolumn{1}{r}{} &   & & \texttt{acf}                &  1 &     & \texttt{int.per}         & 56 \\
             &              & \multicolumn{1}{r}{} &   & &                             &    &     & \texttt{cor}             & 49 \\\hline
         
Absolute     & \underline{\texttt{cdm}}         & 1&18 & & \texttt{pdc}                & 36 &     & \underline{\texttt{cdm}} & 64 \\
Shifting     & \texttt{ncd}                     & 1&60 & & \underline{\texttt{cdm}}    & 19 &     & = \texttt{pdc}           & 64 \\
             & \texttt{cor}                     & 2&20 & & \texttt{spec.llr}           & 13 &     & \texttt{ncd}             & 50 \\
             &              & \multicolumn{1}{r}{} &   & & 4 more                 & $\leq5$ &\ ea.& \texttt{int.per}         & 49 \\
             &              & \multicolumn{1}{r}{} &   & &                             &    &     & \texttt{per}             & 45 \\\hline

Absolute     & \underline{\texttt{cdm}}         & 1&26 & & \texttt{dtw}                & 54 &     & \texttt{dtw}             & 67 \\
Noise        & \texttt{cid}                     & 1&60 & & \underline{\texttt{cdm}}    & 13 &     & \texttt{cid}             & 62 \\
             & \texttt{dtw}                     & 1&63 & & \texttt{cid}                &  5 &     & \underline{\texttt{cdm}} & 50 \\
  & 4 more & \multicolumn{1}{@{}r@{}}{$<2$\phantom{.}} &&& 5 more                 & $\leq4$ &\ ea.& \texttt{infnorm}         & 49 \\
             &              & \multicolumn{1}{r}{} &   & &                             &    &     & \texttt{wav}             & 43 \\\hline

Absolute     & \underline{\texttt{cdm}}         & 1&33 & & \underline{\texttt{cdm}}    & 49 &     & \underline{\texttt{cdm}} & 81 \\
All          & \texttt{ncd}                     & 2&09 & & \texttt{int.per}            & 26 &     & \texttt{ncd}             & 79 \\
             & \texttt{cor}                     & 2&20 & & \texttt{acf}                &  4 &     & \texttt{int.per}         & 59 \\
             &              & \multicolumn{1}{r}{} &   & & \texttt{pdc}, \texttt{pacf} &  1 &\ ea.& \texttt{cor}             & 57 \\
             &              & \multicolumn{1}{r}{} &   & &                             &    &     & \texttt{ccor}            & 54 \\
             &              & \multicolumn{1}{r}{} &   & &                             &    &     & \texttt{pacf}            & 44 \\\hline

Percentage   & \texttt{ncd}                     & 0&85 & & \texttt{ncd}                & 63 &     & \texttt{ncd}             & 81 \\
None         & \underline{\texttt{cdm}}         & 0&95 & & \texttt{dtw}                & 13 &     & \texttt{dtw}             & 73 \\
             & \texttt{dtw}                     & 1&19 & & \texttt{pdc}                &  5 &     & \underline{\texttt{cdm}} & 67 \\
             &              & \multicolumn{1}{r}{} &   &(& \underline{\texttt{cdm}}    &  0 & )   \\\hline

Percentage   & \underline{\texttt{cdm}}         & 1&11 & & \texttt{dtw}                & 33 &     & \underline{\texttt{cdm}} & 78 \\
Noise        & \texttt{ncd}                     & 1&41 & & \underline{\texttt{cdm}}    & 20 &     & \texttt{dtw}             & 69 \\
             & \texttt{dtw}                     & 1&50 & & \texttt{per}, \texttt{pacf} & 11 &\ ea.& \texttt{ncd}             & 55 \\
  & 6 more & \multicolumn{1}{@{}r@{}}{$<2$\phantom{.}} &&& 3 more                      &  2 &\ ea.& \texttt{cid}             & 41 \\\hline 

\end{tabular}
\end{center}

\caption{Summary of experiment described in \autoref{s:exp}.
Refer to the main text for details. In the last column, only listed are measures that rank in the top five spots more than 40 times, i.e.\ for more than half of the possible combinations of warp factor and delay.\label{t:standings}}
\end{table}

\end{document}